\ifx\XeTeXversion\undefined\pdfoutput=1\fi
\documentclass[10pt,twocolumn]{article}
\usepackage[a4paper,margin=1in,bottom=27mm,includefoot,columnsep=6mm]{geometry}
\usepackage[T1]{fontenc}
\usepackage{newtxtext,newtxmath}
\usepackage{amsmath,cite,xurl}
\usepackage{graphicx}
\usepackage{booktabs,array}
\usepackage[hidelinks,pdftitle={JIT-MPE: Just-in-Time Multi-Pitch Estimation under Output Deadlines},pdfauthor={Yoshiyuki Kobayashi}]{hyperref}

\newcommand{\lup}{$\blacktriangle$}
\newcommand{\lne}{$\circ$}
\newcommand{\ldn}{$\blacktriangledown$}
\newcommand{\cv}[3]{\begin{tabular}[t]{@{}c@{}}#1\\[-1pt]{\scriptsize #2}\\[-2pt]{\scriptsize #3}\end{tabular}}

\newcommand{\jitz}{JIT-MPE$_{d=0}$}

\title{JIT-MPE: Just-in-Time Multi-Pitch Estimation\\under Output Deadlines}
\author{Yoshiyuki Kobayashi\\Frieve}
\date{}



\begin{document}

\twocolumn[
\begin{@twocolumnfalse}
\maketitle
\begin{abstract}
Live pitch displays, score followers and accompaniment systems must know which pitches are sounding within tens of milliseconds, but instrument-agnostic multi-pitch estimation (MPE) has mainly been designed and evaluated offline. We present JIT-MPE, a causal MPE model of 86k parameters for mixed ensembles. It gives an immediate estimate of each 20\,ms frame and re-estimates the same frame as 20, 40, 80 and 160\,ms of further audio arrive, so that a live display can show pitches at once and then refine them. We score each output only at a deadline by which it can be computed from the audio that has arrived, on five development sets and four held-out sets evaluated only after all design choices were frozen. At a 50\,ms deadline, which reads the update made with 40\,ms of lookahead, JIT-MPE is significantly more accurate than five of six published offline MPE systems in 38 of 45 comparisons (15 of 20 on the held-out sets) and less accurate in only two, although these systems read the whole recording and all but one have 6 to 170 times as many parameters; the sixth, supervised on URMP, is higher on four of the nine sets. The deadline costs little: on the held-out sets, JIT-MPE at 50\,ms is within $-1.27$ to $+0.60$ points of recording-level macro F1 of its own final output with 160\,ms of lookahead and within $-2.08$ to $+0.79$ points of a bidirectional model of similar size. Waiting pays off where the immediate estimate is weakest: the update at 50\,ms adds 1.7--3.6 points over it on the development data and three of the four held-out sets, and raises the recall of reference frames in the first 20\,ms after a note onset from 9.8\% to 38.4\% on the held-out AAM set. The slowest measured frame takes at most 4.15\,ms on one CPU thread. Code and trained weights are available at \url{https://github.com/Frieve-A/jit-mpe}.
\end{abstract}
\vspace{1.5em}
\end{@twocolumnfalse}
]

\section{Introduction}\label{sec:intro}

Multi-pitch estimation (MPE) reports which pitches sound in each short frame of a recording. When MPE serves offline transcription, the estimator can read the whole recording before it answers. Interactive uses cannot wait: a live pitch display, a score follower or an accompaniment system must decide from the audio that has already arrived, within tens of milliseconds. A recent piano score follower, for example, reports competitive transcription accuracy at a latency of 174\,ms~\cite{peter2025}. Uses that need about 10\,ms, such as the response of a played instrument~\cite{wessel2002problems}, are outside our scope.

Two lines of work each cover half of this setting. Instrument-agnostic MPE models, which accept any combination of instruments, have mainly been designed and evaluated offline: they read the whole recording, and their frame-wise scores do not depend on when each answer becomes available~\cite{weiss2022comparing,bittner2022lightweight,cwitkowitz2024timbretrap,cwitkowitz2024ssmpe}. Low-latency transcription, conversely, has mainly targeted solo instruments, above all the piano~\cite{dessein2010realtime,kong2021high,kwon2024efficient,kusaka2024mobileamt,hu2025minimum}. How much accuracy MPE for mixed ensembles loses when it must answer by a deadline has therefore, to our knowledge, not been measured.

We present JIT-MPE (just-in-time multi-pitch estimation), a causal model of 86k parameters designed for such uses (Figure~\ref{fig:timing}). A live display should show a pitch as soon as it sounds, yet the first milliseconds of a note rarely suffice to resolve its pitch. JIT-MPE therefore gives two kinds of output from one model. It estimates each 20\,ms frame as soon as the audio up to that frame has arrived, and its delay heads re-estimate the same frame 20, 40, 80 and 160\,ms later from the additional audio. A display can draw the immediate estimate at once and rewrite its recent history with each update, like a piano roll that settles just behind the play head, keeping the final estimate at 160\,ms. A model with a single fixed lookahead provides only one of these stages.

To measure what each stage is worth, we score JIT-MPE at each query time only on the output it can compute from the audio available by an output deadline $D$, as in latency-aware evaluation of speech recognition, streaming perception and simultaneous translation (Section~\ref{sec:related}). Each deadline then reads one stage of the update, so the scores show how accurate an early estimate is and what waiting longer improves. Two references measure the cost of the deadline: the final output of the same model isolates the effect of shortening the lookahead, and a bidirectional counterpart of similar size, which we train on the same data and which reads the whole recording, gives a reference without a deadline. Six published offline MPE systems and two published multi-instrument transcription systems, grouped by their training data, place the absolute accuracy.

We develop on chamber, chorale and orchestral recordings (URMP, Bach10, MusicNet, TRIOS and PHENICX-Anechoic). After freezing every design choice, we evaluate on three held-out datasets: AAM, synthetic multitrack songs with many instruments; ChoraleBricks, wind-ensemble chorales mixed from separately recorded parts; and BSED, Beethoven symphony excerpts in four real orchestral performances and one synthetic rendering, which we report separately as BSED-real and BSED-synthetic. With five development sets, this gives nine evaluation sets.

The results show several strengths of this design.
\begin{itemize}
\item \emph{Higher accuracy than larger offline systems.} At $D = 50$\,ms, which reads the update made with 40\,ms of lookahead, JIT-MPE is significantly higher than five of six published offline MPE systems, which read the whole recording, in 38 of 45 comparisons on the nine evaluation sets (15 of 20 on the held-out sets) and lower only in two, both on BSED-real. All six except Basic Pitch have 6 to 170 times as many parameters. The sixth, SS-MPE v0.2.0, is supervised on URMP and is higher on the three sets with continuous-f0 truth (URMP, Bach10-mf0-synth and ChoraleBricks) and on BSED-real. Of two published transcription systems, about 550 and 16,000 times larger, MuScriptor is higher on AAM and BSED-real.
\item \emph{Little cost of the deadline.} On the four held-out sets, JIT-MPE at 50\,ms is within $-1.27$ to $+0.60$ points of recording-level macro F1 of its own final output and within $-2.08$ to $+0.79$ points of its bidirectional counterpart, which is significantly higher only on AAM.
\item \emph{Updates that pay off where they are needed.} The update read at 50\,ms adds 1.7--3.6 points over the immediate estimate on the development data and on three of the four held-out sets, most of the gap to the final output. The updates help most just after note onsets: on AAM, the recall of reference frames in the first 20\,ms after an onset rises from 9.8\% for the immediate estimate to 38.4\% at 50\,ms and 62.3\% at 160\,ms.
\item \emph{An immediate estimate that holds its own.} Read through its immediate ($d = 0$) head alone, JIT-MPE is on no set significantly lower than Harmonic CRNN, a single-head model with the same harmonic front end (Section~\ref{sec:variants}).
\item \emph{Small and fast.} JIT-MPE needs 0.68\,GMAC per second of audio, and on one CPU thread its slowest measured 20\,ms frame takes at most 4.15\,ms.
\end{itemize}

The contributions are:
\begin{itemize}
\item JIT-MPE, a causal MPE model of 86k parameters that gives an immediate estimate of each frame and re-estimates it as more audio arrives, so that one model serves every deadline, released with code and trained weights (Section~\ref{sec:models}).
\item A deadline-constrained evaluation of MPE, which scores each output only at a deadline by which it can be computed and thereby measures how accurate each stage of the update is (Section~\ref{sec:protocol}).
\item A measurement, on development data and three held-out datasets, of the accuracy kept at a short deadline and of what waiting improves, above all the frames just after note onsets (Sections~\ref{sec:cost}, \ref{sec:onset} and~\ref{sec:ablation}).
\item A comparison with eight published systems grouped by training data, which places the absolute accuracy (Section~\ref{sec:published}).
\end{itemize}

\section{Related Work}\label{sec:related}

\textbf{Published offline MPE systems.} The following systems estimate multiple pitches for arbitrary instrument combinations and are designed and evaluated on full recordings. Deep salience models estimate pitch salience from a harmonic constant-Q representation~\cite{bittner2017deep}. Wei{\ss} and Peeters compare deep MPE architectures and show that results depend on the evaluation strategy~\cite{weiss2022comparing}; we use their released DRCNN and PUnet XL models (WP-DRCNN and WP-PUnet XL) as references. Basic Pitch is a lightweight model for instrument-agnostic note transcription and MPE~\cite{bittner2022lightweight}. Timbre-Trap targets instrument-agnostic transcription with little labelled data~\cite{cwitkowitz2024timbretrap}. SS-MPE learns MPE without labels~\cite{cwitkowitz2024ssmpe}, and later work studies overfitting and degeneration in this setting~\cite{cwitkowitz2025overfitting}. In piano transcription, HPPNet builds harmonic structure into the model with harmonically dilated convolutions~\cite{wei2022hppnet}.

\textbf{Multi-instrument transcription.} MT3 and YourMT3+ transcribe multi-instrument audio into note sequences with Transformer models~\cite{gardner2022mt3,chang2024yourmt3}, and MuScriptor is an open multi-instrument transcription model~\cite{rouard2026muscriptor}. They output notes rather than frame-level pitch; we include YourMT3+ and MuScriptor as reference points.

\textbf{Low-latency transcription.} Piano transcription has been studied at high temporal resolution~\cite{kong2021high} and in real time~\cite{dessein2010realtime,kwon2024efficient,kusaka2024mobileamt}. Hu et al.\ adapt piano transcription systems for minimum latency and report how accuracy changes~\cite{hu2025minimum}; this is the closest prior study we know of, but it covers the piano only. Two concurrent papers study the accuracy--latency trade-off for piano and guitar with a band-split RNN that reduces the dependence on future input~\cite{kusaka2026lowlatency}, and real-time violin transcription for score following~\cite{lee2026stradi}. We instead study instrument-agnostic MPE on mixed ensembles at fixed deadlines.

\textbf{Latency-aware evaluation.} Baumann et al.\ score incremental speech recognition by the accuracy of the hypotheses available at fixed lags~\cite{baumann2009assessing}. Streaming perception scores only the outputs available when the world is queried~\cite{li2020streaming}, and SimulEval measures the quality and latency of simultaneous translation~\cite{ma2020simuleval}. We apply the same idea to MPE. In streaming speech recognition, one model can serve several latencies or limit its right context~\cite{yu2021dualmode,shi2021emformer}; the delay heads of JIT-MPE follow this principle.

\section{Deadline-Constrained Evaluation}\label{sec:protocol}

\begin{figure}[t]
  \centering
  \includegraphics[width=\columnwidth]{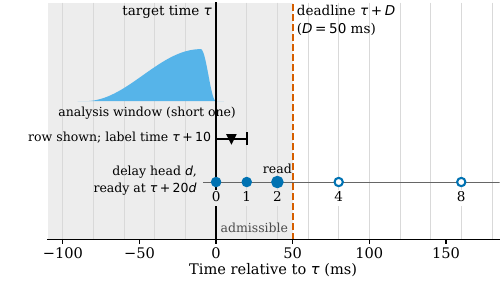}
  \caption{Timing of one output row of JIT-MPE. The analysis window (the shorter of the two is shown) ends at the target time $\tau$, so the $d=0$ head needs no audio after $\tau$. The bracket marks the 20\,ms during which the row is shown; its training label refers to the middle, $\tau + 10$\,ms (triangle). Delay head $d$ re-estimates the same row from the audio up to $\tau + 20d$\,ms. With a deadline of $D = 50$\,ms, heads 0, 1 and 2 are ready in time (filled markers), and head 2, the one selected for this deadline (Section~\ref{sec:calib}), is read (large marker); heads 4 and 8 (hollow markers) are not ready in time.}
  \label{fig:timing}
\end{figure}

We score a causal system only on what it can output by a fixed deadline.

\subsection{Frames and delay heads}

Our models run at a hop of 20\,ms. Output row $t$ has the target time $\tau_t = (t+1) \cdot 20$\,ms, and its analysis windows end at $\tau_t$ (Figure~\ref{fig:timing}). Its training label refers to the label time $\tau_t + 10$\,ms, the middle of the 20\,ms during which the clock below shows the row. A delay head $d$ of row $t$ reads the state of the model's GRU (Section~\ref{sec:jit}) at row $t+d$ and therefore needs the audio up to $\tau_t + 20d$\,ms. JIT-MPE has heads $d \in \{0,1,2,4,8\}$, that is, waits of 0, 20, 40, 80 and 160\,ms. For each row they form a sequence of updates: the $d = 0$ head gives the immediate estimate, which uses no audio after $\tau_t$ (its computation still takes time), heads 1, 2 and 4 update it 20, 40 and 80\,ms later, and head 8 gives the final estimate. A deadline selects one stage of this sequence; at $D = 50$\,ms it is the update of head 2, made with 40\,ms of lookahead (Section~\ref{sec:calib}), not the immediate estimate.

\subsection{Output clock}

A query at time $q$ asks which pitches sound at $q$. Under a deadline $D$, the answer may use only the audio up to $q + D$. Our clock answers with row $r$, the latest row whose target time is not after $q$ ($\tau_r \le q < \tau_r + 20$\,ms), read through the head $d_D$ selected for $D$ (Section~\ref{sec:calib}). Heads are selected only among those with $20 d_D \le D$, so the audio up to $\tau_r + 20 d_D \le q + D$ that the output needs has always arrived. We call this the \emph{stale-hold} clock: like a display, it keeps showing the latest row until the next one is due, and the row's label time lies within 10\,ms of $q$.

The deadline $D$ bounds the audio an output may use. It does not include computation time or the delays of the audio interface. At $D = 50$\,ms, the selected head needs audio only up to $q + 40$\,ms, which leaves at least 10\,ms for computation; Section~\ref{sec:compute} shows that the computation fits into this slack. At the deadlines that are multiples of 20\,ms, the needed audio can arrive exactly at $q + D$, so there $D$ bounds the lookahead only. Since no head waits 50\,ms, $D = 50$\,ms and $D = 40$\,ms read the same head and give the same scores.

\subsection{Deadlines}

We evaluate deadlines of 0, 20, 40, 50 and 60\,ms, every 20\,ms from 80 to 200\,ms, and full length. The tables report $D = 50$\,ms, and Figure~\ref{fig:curves} shows all deadlines. At full length, JIT-MPE uses its selected head $d = 8$, that is, 160\,ms of lookahead, whereas bidirectional models and the published systems read the whole recording.

\subsection{Metrics}\label{sec:metrics}

The truth is queried on the time grid of each dataset: every 10\,ms from 0, except URMP and Bach10, whose f0 annotations lie at $23 + 10k$\,ms. Full-length outputs take the output frame nearest to each query time (for our models, by target time; an exact tie goes to the earlier frame) and are empty outside their range. On the grids that start at 0, this is the stale-hold row; on URMP and Bach10 it is often the next row, whose label time is further from $q$, which slightly lowers the full-length scores of our models there (Appendix~\ref{s:caveats}).

Our models output 88-key semitone presence; for continuous-pitch scoring, each active semitone is refined in 20-cent steps from the spectrum of its own target row, which adds no lookahead. The score is recording-level macro F1, the mean of the per-recording F1.

Each dataset has a \emph{primary truth}: continuous f0 where the dataset provides it (URMP, Bach10-mf0-synth and ChoraleBricks), scored with continuous-f0 F1, and nominal notes otherwise, scored with semitone F1. Continuous-f0 F1 matches estimated and reference pitches one-to-one within 50 cents. Semitone F1 compares sets of MIDI notes. To handle recordings whose tuning departs from A440, we estimate for each recording the offset
\[
c = 100 \cdot \mathrm{median}\big(m - \mathrm{round}(m)\big)
\]
in cents over all voiced reference values $m = \mathrm{MIDI}(f_0)$, and map every reference and estimated pitch $m$ to the note $\lfloor m - c/100 + 0.5 \rfloor$; without a continuous-f0 reference, $c = 0$. Semitone truth uses note labels where they exist and the rounded f0 otherwise.

The two truths can differ: on the nine URMP recordings of note-19 (Section~\ref{sec:setup}), the rounded note truth and the rounded f0 truth disagree on 9.18\% of 66,608 active frames. We therefore base the comparison on the primary truth of each dataset and also report the semitone metric on ChoraleBricks as a uniform check. Appendix~\ref{s:eval} gives the full scoring procedure.

\subsection{Statistics}

Each of our models is trained with three seeds, and we report the mean over seeds. Confidence intervals come from a paired bootstrap with 10,000 draws that resamples work clusters (pieces, songs or symphonies) within each dataset; within each draw we compute every system and every paired difference on the same resampled recordings. We report percentile 95\% CIs. They reflect the variation over recordings, not over training runs; across seeds, the per-set F1 of JIT-MPE at 50\,ms varies with a standard deviation of at most 0.69 points, and that of Harmonic BiCRNN, its bidirectional counterpart (Section~\ref{sec:variants}), of at most 1.40.

We read a paired difference as \emph{above} (significantly better) if its CI lies entirely above zero, \emph{below} (significantly worse) if it lies entirely below zero, and \emph{not significantly different} otherwise; the last reading is not a claim of equivalence. We make many such comparisons without correcting for multiplicity, so a single reading near the boundary should not carry a conclusion. The comparison tables never pool datasets. The development aggregate All-29, the 29 recordings of URMP F0s, Bach10-mf0-synth and MusicNet described in Section~\ref{sec:setup}, used in Tables~\ref{tab:variants}, \ref{tab:cost} and~\ref{tab:ladder} and Figure~\ref{fig:curves}, summarizes development results only.

\section{Models and Training}\label{sec:models}

\subsection{JIT-MPE}\label{sec:jit}

Each 20\,ms frame is described by two analysis windows of 85\,ms and 341\,ms that end at the target time. The magnitude spectrum of each window is projected onto a log-frequency axis with three bins per semitone (300 bins, MIDI 20.67--120.33), log-compressed and standardized. The windows are asymmetric, with a long rise and a short fall of 10\,ms (short window) and 20\,ms (long window), following Hu et al., who use such windows to cut the delay of the front end~\cite[Sec.~4.2]{hu2025minimum}.

A convolutional \emph{harmonic stem} of eight layers (24 channels, $3 \times 7$ kernels, frequency dilations from 1 to 64) sees $\pm 384$ bins, more than the whole pitch range. Each kernel is causal over three frames, so the stem sees the last 17 frames. A harmonic gather then reads, for each of the 88 notes, the stem features at 18 harmonic and subharmonic positions of that note and forms 64 features per note. An interval aggregation combines the features of each note with those of notes at fixed intervals (unison, seconds, fourths and fifths, octaves, twelfths and double octaves) through a rank-16 residual. A GRU with 64 units, shared across notes, runs over time for each note. Linear presence heads for $d \in \{0,1,2,4,8\}$ read the GRU state of row $t+d$ to estimate row $t$; all heads are trained jointly. Appendix~\ref{s:model} gives the full definition.

\subsection{Model variants}\label{sec:variants}

\begin{table}[t]
\centering
\small
\setlength{\tabcolsep}{2.5pt}
\begin{tabular}{lrrrrr}
\toprule
 & Radius & \multicolumn{2}{c}{Parameters} & \multicolumn{2}{c}{F1, All-29 (\%)} \\
\cmidrule(lr){3-4}\cmidrule(lr){5-6}
Model & (st) & causal & bidir. & causal & bidir. \\
\midrule
Dilated CRNN & 47 & 95,767 & 93,936 & 75.93 & 79.84 \\
Harmonic CRNN & 128 & 85,813 & 83,729 & 76.47 & 80.46 \\
JIT-MPE & 128 & 86,425 & --- & 80.55 & --- \\
\bottomrule
\end{tabular}
\caption{Our models, from the ablation's starting point to JIT-MPE (Section~\ref{sec:variants}); a bidirectional counterpart (BiCRNN) of each of the first two is trained on the same data. Radius: frequency reach of the convolutional front end in semitones. Parameters exclude the onset output layer used only for an auxiliary training loss. F1: development aggregate All-29, full-length outputs, primary truth, mean of three seeds.}
\label{tab:variants}
\end{table}

To explain the accuracy of JIT-MPE and to measure the cost of the deadline, we train further models of our own (Table~\ref{tab:variants} and Appendix~\ref{s:model}). None of them is a published system; all are trained on the same data with the same recipe as JIT-MPE.

\textbf{Causal variants.} \emph{Dilated CRNN} is a convolutional recurrent network: three $3 \times 7$ convolutions with 32 channels and frequency dilations 1, 6 and 36, a strided projection to 47 features per note and a 47-unit GRU shared across notes. Its front end sees $\pm 47$ semitones, enough to relate a note to its overtones up to about the 15th harmonic. A front end must see these overtones; harmonic input representations~\cite{bittner2017deep} and harmonically dilated convolutions~\cite{wei2022hppnet} build this in, and the same CRNN without dilation, which sees only $\pm 7$ semitones, is 9 to 18 points lower on every set (Plain CRNN, Appendix~\ref{s:model}). We therefore start the ablation from Dilated CRNN. \emph{Harmonic CRNN} replaces this front end with the harmonic stem, gather, interval aggregation and GRU of JIT-MPE, but uses symmetric Hann windows of the same lengths and a single head $d = 0$. JIT-MPE adds the asymmetric windows and the delay heads. We also read JIT-MPE through its $d = 0$ head alone and call this \jitz; it is the same trained model, not a separate one.

These models form a ladder of three steps: the harmonic front end, that is, stem, gather and interval aggregation together (Harmonic $-$ Dilated CRNN); the window together with multi-head training (\jitz\ $-$ Harmonic CRNN); and the later read of the delay heads (JIT-MPE $-$ \jitz). The first two steps compare separately trained models, and each bundles several changes. The last step compares two read-outs of one trained model and measures what the additional audio is worth to JIT-MPE.

\textbf{Bidirectional counterparts.} \emph{Dilated BiCRNN} and \emph{Harmonic BiCRNN} keep the frequency layout of the corresponding causal model but use time-centred convolutions and a bidirectional GRU, with 26 and 37 units per direction, so that their sizes match. They run on the whole recording at once. Harmonic BiCRNN is the counterpart of JIT-MPE. The two differ in causality and also in the window, the delay heads and the recurrent layer, whose capacity Harmonic BiCRNN splits into two directions. We therefore call JIT-MPE $-$ Harmonic BiCRNN the \emph{causal--bidirectional gap} and do not read it as a pure effect of causality.

\subsection{Training}\label{sec:training}

All our models use the same 767 training clips from public datasets (Table~\ref{tab:clips} in Appendix~\ref{s:data}): GuitarSet~\cite{xi2018guitarset}, Slakh2100-redux~\cite{manilow2019slakh}, MusicNet~\cite{thickstun2017musicnet}, Winterreise~\cite{weiss2021winterreise}, Cantoria~\cite{cantoria}, Dagstuhl ChoirSet~\cite{rosenzweig2020dagstuhl}, vocadito~\cite{bittner2021vocadito} and Groove~\cite{gillick2019groove}. The Groove clips contain drums only and serve as negatives. Labels are 88-key semitone presence derived from note annotations or, for the vocal datasets, from f0 annotations; uncertain labels are excluded from the loss.

Training has two stages. Stage 1 trains from random initialization on random labelled crops of 3.84\,s for 10,400 steps with the binary cross-entropy of each presence head, averaged over heads. Stage 2 continues for 3,000 steps at a lower learning rate. It mixes drum and vocal excerpts into the crops, adds synthetic low-register tones to crops that lack low notes, applies random compression and level changes, and adds an auxiliary onset loss. Appendix~\ref{s:data} gives the labels and the augmentation.

\subsection{Calibration and head selection}\label{sec:calib}

After training, each model and head receives an additive bias on its output logits, one global value and three register-specific values, fitted on calibration clips held out from the training datasets. For each deadline, one head is then chosen on the stale-hold clock on a separate selection set of 187 clips from four training datasets, disjoint by work from the training clips; no evaluation set is used. A head is a candidate for deadline $D$ only if it is ready in time ($20d \le D$\,ms). For JIT-MPE, the selected head is $d = 1$ at 20\,ms, $d = 2$ from 40 to 60\,ms, $d = 4$ from 80 to 140\,ms and $d = 8$ from 160\,ms and at full length, for all three seeds. Appendix~\ref{s:calib} gives the details.

\section{Experimental Setup}\label{sec:setup}

\subsection{Development data}

The continuous-f0 comparison uses URMP F0s~\cite{li2019urmp} (9 recordings from 7 pieces) and Bach10-mf0-synth~\cite{duan2010bach10,salamon2017analysis} (10 recordings of Bach10, resynthesized so that their f0 annotations are exact). The semitone comparison uses note-19, the 19 note-labelled recordings of MusicNet (10) and URMP (9) from 17 pieces; TRIOS~\cite{fritsch2013trios} (5 recordings); and PHENICX-Anechoic~\cite{miron2016phenicx} (4 recordings). We looked at TRIOS and PHENICX results before freezing the models, so we count them as development data. All-29 combines URMP F0s, Bach10-mf0-synth and the 10 MusicNet recordings, each on its primary truth.

\subsection{Held-out data}

After freezing all models, calibrations and head selections, we evaluated three held-out datasets; nothing was changed after their scores were seen.
\begin{itemize}
\item \textbf{AAM}~\cite{ostermann2023aam}: 100 songs drawn at random from the first 2,000 of its 3,000 songs, which we had reserved, each a 30\,s mono mix of all instruments. Drums are excluded from the truth. The songs are unseen, but synthetic multitrack music is part of our training domain through Slakh.
\item \textbf{ChoraleBricks}~\cite{balke2025choralebricks}: 10 wind-ensemble chorales, each mixed from its separately recorded parts (193 stems of 13 instruments in total). It provides both continuous f0 and notes; its primary truth is f0.
\item \textbf{BSED}~\cite{berendes2026bsed}: 20 excerpts from 9 Beethoven symphonies, each in four real performances and one synthetic rendering, scored against note-level alignments. We report the 80 real performances (BSED-real) and the 20 synthetic renderings (BSED-synthetic) separately.
\end{itemize}
Work clusters are songs for AAM and ChoraleBricks and symphonies for BSED. Appendix~\ref{s:eval} gives the construction of each set.

\textbf{Overlap with our training data.} No evaluation recording is in our training data; our training data are public and recording-disjoint from all evaluation sets. MusicNet is in our training pool, so note-19 (through its MusicNet recordings) and BSED (through the Beethoven repertoire of MusicNet) are partly in-domain for our models. We have not checked whether works in Bach10, TRIOS, PHENICX-Anechoic and ChoraleBricks also occur in our training data.

\begin{table*}[t]
\centering
{\footnotesize\setlength{\tabcolsep}{5pt}\begin{tabular}{lccccc}
\toprule
 & Development & \multicolumn{4}{c}{Held-out} \\
\cmidrule(lr){2-2}\cmidrule(lr){3-6}
System & All-29 & AAM & ChoraleBricks & BSED-real & BSED-syn. \\
\midrule
JIT-MPE, $D = 50$\,ms & \textbf{80.83} & 82.58 & \textbf{77.96} & 64.52 & 77.01 \\[6pt]
JIT-MPE, full length (160\,ms lookahead) & \cv{80.55}{$+0.28$\,\lne}{[$-0.001$, 0.58]} & \cv{83.85}{$-1.27$\,\ldn}{[$-1.40$, $-1.14$]} & \cv{77.36}{$+0.60$\,\lup}{[0.47, 0.71]} & \cv{\textbf{65.49}}{$-0.97$\,\ldn}{[$-1.27$, $-0.72$]} & \cv{\textbf{78.28}}{$-1.27$\,\ldn}{[$-1.70$, $-0.92$]} \\[6pt]
Harmonic BiCRNN, offline (whole recording) & \cv{80.46}{$+0.38$\,\lne}{[$-0.03$, 0.76]} & \cv{\textbf{84.67}}{$-2.08$\,\ldn}{[$-2.55$, $-1.62$]} & \cv{77.17}{$+0.79$\,\lup}{[0.08, 1.49]} & \cv{65.04}{$-0.52$\,\lne}{[$-1.24$, 0.27]} & \cv{77.50}{$-0.50$\,\lne}{[$-1.03$, 0.06]} \\
\bottomrule
\end{tabular}}
\caption{Accuracy at the deadline, on the primary truth of each set; all rows are our models. Each cell gives the recording-level macro F1 (\%) in normal type, the highest in each column in bold, and below it in small type the paired difference $\Delta$ (points) with its 95\% bootstrap CI. $\Delta$ = JIT-MPE at $D = 50$\,ms $-$ the system of the row, so a negative value is the cost of the deadline. JIT-MPE at 50\,ms is read on the stale-hold clock; its full-length output and Harmonic BiCRNN take the nearest output frame (Section~\ref{sec:metrics}). \lup\ above, \lne\ not significantly different, \ldn\ below. Bounds that would round to zero are given to their first nonzero decimal.}
\label{tab:cost}
\end{table*}

\subsection{Published reference systems}

We compare with six published offline MPE systems, five models with SS-MPE in two releases, and two published transcription systems, all with the checkpoints released by their authors (Appendix~\ref{s:eval}). Their training data matter for interpretation, because several are trained on our evaluation datasets. Their size and computation are given in Table~\ref{tab:footprint} (Section~\ref{sec:compute}).
\begin{itemize}
\item \textbf{SS-MPE v0.1.0}~\cite{cwitkowitz2024ssmpe} is self-supervised on NSynth~\cite{engel2017nsynth}.
\item \textbf{SS-MPE v0.2.0} is supervised on URMP, and its released checkpoint was chosen on the URMP F0 recordings that we also evaluate.
\item \textbf{Basic Pitch}~\cite{bittner2022lightweight} is trained on MAESTRO~\cite{hawthorne2019maestro}, Slakh, GuitarSet, iKala~\cite{chan2015ikala} and MedleyDB~\cite{bittner2014medleydb}.
\item \textbf{Timbre-Trap}~\cite{cwitkowitz2024timbretrap} is trained on URMP, and its checkpoint was selected on a nine-recording URMP subset.
\item \textbf{WP-DRCNN} (size L) and \textbf{WP-PUnet XL}~\cite{weiss2022comparing} are trained on MusicNet.
\item \textbf{YourMT3+}~\cite{chang2024yourmt3} is trained on data that include URMP and MusicNet.
\item \textbf{MuScriptor}~\cite{rouard2026muscriptor} is trained on audio synthesized from public and commercial MIDI files and on internal real recordings; its authors evaluate on Bach10 and PHENICX-Anechoic as data outside its training set. We use guidance strength 2 (CFG2).
\end{itemize}
All thresholds are author defaults, and none is tuned on our evaluation data. Our models, in contrast, use biases fitted on calibration clips held out from their training datasets (Section~\ref{sec:calib}) and our pitch refinement; the comparison is therefore one of complete pipelines as released, not of model architectures alone. The two transcription systems output notes, not frame-level pitch, so we score their native note output as frame-level pitch, a use outside their purpose; these scores do not measure their quality as transcribers.

\section{Results}\label{sec:results}

\subsection{Cost of the deadline}\label{sec:cost}

Table~\ref{tab:cost} measures the cost of the deadline in two ways: against the full-length output of JIT-MPE itself, which looks 160\,ms ahead and so isolates the effect of shortening the lookahead, and against Harmonic BiCRNN, a reference point of similar size trained on the same data that sees the whole recording. Within JIT-MPE, answering at 50\,ms instead of with 160\,ms of lookahead costs 1.27 points on AAM, 0.97 on BSED-real and 1.27 on BSED-synthetic, and gains 0.60 on ChoraleBricks. On the development data the difference is not significant. Against Harmonic BiCRNN, JIT-MPE at 50\,ms is significantly lower only on AAM ($-2.08$) and significantly higher on ChoraleBricks ($+0.79$); on the development data and on both BSED sets the difference is not significant. When JIT-MPE also reads its full-length output, it is within $-0.81$ to $+0.78$ points of Harmonic BiCRNN. On AAM, therefore, about 1.3 of the 2.1 points at 50\,ms are due to the deadline, and about 0.8 remain between JIT-MPE with 160\,ms of lookahead and Harmonic BiCRNN, which reads the whole recording.

\begin{figure*}[t]
  \centering
  \includegraphics[width=\textwidth]{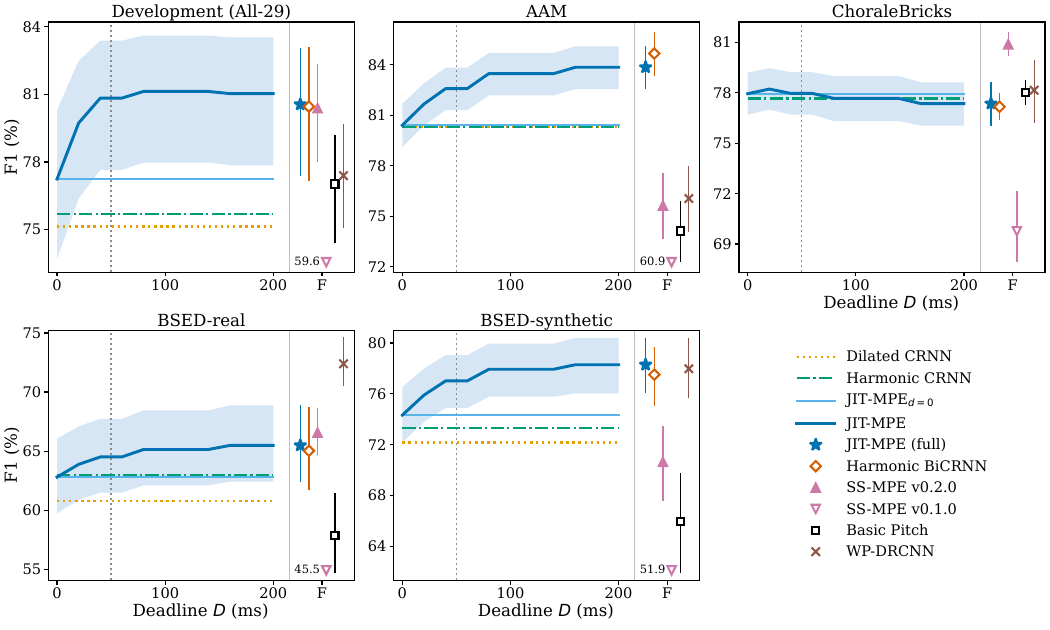}
  \caption{Recording-level macro F1 on the primary truth against the output deadline $D$ (0--200\,ms), which bounds the audio each answer may use. Shown for the development aggregate All-29 and the four held-out sets. Lines: causal models on the stale-hold clock; the single-head models and \jitz\ read $d = 0$ at every deadline and give flat lines. The dashed vertical line marks $D = 50$\,ms. Shaded band: 95\% CI of JIT-MPE. Markers at F: full-length outputs with 95\% CIs. Points below the plotted range are drawn at the bottom of the panel with their values. Line styles and marker shapes distinguish the series also in greyscale.}
  \label{fig:curves}
\end{figure*}

Figure~\ref{fig:curves} shows how the accuracy depends on the lookahead that the deadline allows. On the development data, JIT-MPE reaches 77.24\% at 0\,ms, 79.73\% at 20\,ms and 80.83\% from 40 to 60\,ms, and stays at 81.03--81.13\% from 80 to 200\,ms, so waiting longer than 40\,ms gains at most 0.3 points there. On AAM, its F1 rises from 80.41\% at 0\,ms to 82.58\% at 50\,ms, 83.48\% from 80 to 140\,ms and 83.85\% at full length; on BSED-real from 62.81\% to 64.52\% and 65.49\%, and on BSED-synthetic from 74.32\% to 77.01\% and 78.28\%. ChoraleBricks is nearly flat: 77.95\% at 0\,ms, 78.22\% at 20\,ms, 77.96\% at 50\,ms and 77.36\% at full length. The curves change only where a new head becomes available (at 20, 40, 80 and 160\,ms), so their flat segments follow from these discrete delays and do not mark an optimal deadline.

\begin{table*}[t]
\centering
{\footnotesize\setlength{\tabcolsep}{8pt}\begin{tabular}{llccc}
\toprule
 & & URMP F0s & Bach10-mf0-synth & ChoraleBricks f0 \\
System & Training domain & (dev, 9/7) & (dev, 10/10) & (held-out, 10/10) \\
\midrule
\multicolumn{5}{l}{\textit{Ours: causal, output deadline $D = 50$\,ms}} \\
JIT-MPE, $D=50$\,ms & public, recording-disjoint & 84.67 & 91.06 & 77.96 \\
\specialrule{.4pt}{\aboverulesep}{1.5pt}\specialrule{.4pt}{0pt}{\belowrulesep}
\multicolumn{5}{l}{\textit{Published systems: offline, each reads the whole recording}} \\
SS-MPE v0.1.0 & NSynth & \cv{60.81}{$+23.87$\,\lup}{[17.84, 31.52]} & \cv{72.28}{$+18.79$\,\lup}{[17.48, 20.24]} & \cv{69.76}{$+8.20$\,\lup}{[6.37, 9.92]} \\[8pt]
SS-MPE v0.2.0 & URMP (sup.) & \cv{\textbf{88.39$^\dagger$}}{$-3.72$\,\ldn}{[$-5.51$, $-1.87$]} & \cv{\textbf{92.87}}{$-1.80$\,\ldn}{[$-2.20$, $-1.38$]} & \cv{\textbf{80.88}}{$-2.92$\,\ldn}{[$-3.85$, $-1.96$]} \\[8pt]
Basic Pitch & \begin{tabular}[t]{@{}l@{}}MAESTRO, Slakh, GuitarSet,\\iKala, MedleyDB\end{tabular} & \cv{80.86}{$+3.81$\,\lup}{[2.72, 4.65]} & \cv{88.30}{$+2.76$\,\lup}{[1.97, 3.71]} & \cv{78.02}{$-0.06$\,\lne}{[$-0.95$, 0.80]} \\[8pt]
WP-DRCNN & MusicNet & \cv{76.60}{$+8.08$\,\lup}{[4.66, 12.89]} & \cv{82.27}{$+8.80$\,\lup}{[7.78, 10.17]} & \cv{78.15}{$-0.19$\,\lne}{[$-1.64$, 1.24]} \\[8pt]
WP-PUnet XL & MusicNet & \cv{74.74}{$+9.94$\,\lup}{[6.19, 15.10]} & \cv{82.17}{$+8.89$\,\lup}{[7.56, 10.53]} & \cv{73.43}{$+4.53$\,\lup}{[2.85, 6.16]} \\[8pt]
Timbre-Trap & URMP & \cv{73.34$^\dagger$}{$+11.34$\,\lup}{[6.34, 15.17]} & \cv{84.75}{$+6.32$\,\lup}{[5.55, 7.11]} & \cv{54.76}{$+23.20$\,\lup}{[21.31, 25.09]} \\
\bottomrule
\end{tabular}}
\caption{Comparison on the continuous-f0 truth (50-cent one-to-one matching). Each cell gives the recording-level macro F1 (\%) in normal type, the highest in each column in bold, and below it in small type the paired difference $\Delta$ (points) with its 95\% bootstrap CI. $\Delta$ = JIT-MPE $-$ system, so positive values favour JIT-MPE. JIT-MPE is read at $D=50$\,ms on the stale-hold clock; the published systems are scored on their full-length outputs. \lup\ above, \lne\ not significantly different, \ldn\ below. Column headers give recordings/work clusters. $\dagger$: trained or selected on URMP. JIT-MPE emits semitones refined in 20-cent steps; Basic Pitch and the WP models emit semitone frames; SS-MPE and Timbre-Trap emit continuous pitch.}
\label{tab:f0}
\end{table*}

\begin{table*}[t]
\centering
{\footnotesize\setlength{\tabcolsep}{1.8pt}\begin{tabular}{llccccccc}
\toprule
 & & \multicolumn{3}{c}{Development} & \multicolumn{4}{c}{Held-out} \\
\cmidrule(lr){3-5}\cmidrule(lr){6-9}
System & Training domain & note-19$^\ddagger$ & TRIOS & PHENICX & AAM & ChoraleBricks$^\ast$ & BSED-real$^\ddagger$ & BSED-syn.$^\ddagger$ \\
 & & 19/17 & 5/5 & 4/4 & 100/100 & 10/10 & 80/9 & 20/9 \\
\midrule
\multicolumn{9}{l}{\textit{Ours: causal, output deadline $D = 50$\,ms}} \\
JIT-MPE, $D=50$\,ms & public, rec.-disjoint & 77.0 & 77.6 & \textbf{77.2} & 82.6 & 78.9 & 64.5 & 77.0 \\
\specialrule{.4pt}{\aboverulesep}{1.5pt}\specialrule{.4pt}{0pt}{\belowrulesep}
\multicolumn{9}{l}{\textit{Published systems: offline, each reads the whole recording}} \\
SS-MPE v0.1.0 & NSynth & \cv{54.8}{$+22.2$\,\lup}{[18.5, 26.2]} & \cv{58.0}{$+19.6$\,\lup}{[14.3, 25.7]} & \cv{54.0}{$+23.2$\,\lup}{[19.7, 27.2]} & \cv{60.9}{$+21.7$\,\lup}{[19.7, 23.8]} & \cv{71.4}{$+7.5$\,\lup}{[5.6, 9.3]} & \cv{45.5}{$+19.0$\,\lup}{[15.7, 22.7]} & \cv{51.9}{$+25.1$\,\lup}{[21.6, 28.7]} \\[5pt]
SS-MPE v0.2.0 & URMP & \cv{75.0$^\dagger$}{$+2.0$\,\lne}{[$-0.1$, 4.2]} & \cv{72.7}{$+4.9$\,\lne}{[$-1.4$, 10.1]} & \cv{74.3}{$+2.8$\,\lup}{[1.4, 5.0]} & \cv{75.6}{$+7.0$\,\lup}{[5.5, 8.4]} & \cv{\textbf{82.2}}{$-3.3$\,\ldn}{[$-4.3$, $-2.3$]} & \cv{66.6}{$-2.0$\,\ldn}{[$-3.6$, $-0.5$]} & \cv{70.6}{$+6.4$\,\lup}{[5.2, 7.9]} \\[5pt]
Basic Pitch & MAE/Slakh/GS/iK/MDB & \cv{72.6}{$+4.3$\,\lup}{[2.7, 5.8]} & \cv{69.4}{$+8.2$\,\lup}{[4.3, 11.5]} & \cv{74.2}{$+2.9$\,\lup}{[2.0, 3.8]} & \cv{74.1}{$+8.5$\,\lup}{[7.1, 9.9]} & \cv{79.1}{$-0.2$\,\lne}{[$-1.1$, 0.8]} & \cv{57.9}{$+6.6$\,\lup}{[4.9, 8.7]} & \cv{65.9}{$+11.1$\,\lup}{[8.6, 13.5]} \\[5pt]
WP-DRCNN & MusicNet & \cv{76.4$^\dagger$}{$+0.6$\,\lne}{[$-3.1$, 3.7]} & \cv{71.6}{$+6.0$\,\lup}{[2.1, 11.5]} & \cv{73.1}{$+4.1$\,\lup}{[1.2, 7.0]} & \cv{76.1}{$+6.5$\,\lup}{[5.1, 8.0]} & \cv{80.4}{$-1.4$\,\ldn}{[$-2.9$, $-0.001$]} & \cv{\textbf{72.4$^\S$}}{$-7.9$\,\ldn}{[$-9.9$, $-6.0$]} & \cv{\textbf{78.0$^\S$}}{$-1.0$\,\lne}{[$-2.3$, 0.3]} \\[5pt]
WP-PUnet XL & MusicNet & \cv{74.8$^\dagger$}{$+2.2$\,\lne}{[$-1.8$, 5.7]} & \cv{66.7}{$+11.0$\,\lup}{[8.7, 13.3]} & \cv{65.7}{$+11.5$\,\lup}{[11.1, 11.9]} & \cv{72.4}{$+10.2$\,\lup}{[8.6, 11.8]} & \cv{75.6}{$+3.3$\,\lup}{[1.6, 4.9]} & \cv{67.0$^\S$}{$-2.5$\,\ldn}{[$-4.5$, $-0.3$]} & \cv{73.6$^\S$}{$+3.5$\,\lup}{[1.4, 5.6]} \\[5pt]
Timbre-Trap & URMP & \cv{62.1$^\dagger$}{$+14.9$\,\lup}{[11.5, 17.9]} & \cv{58.1}{$+19.5$\,\lup}{[13.9, 24.1]} & \cv{53.7}{$+23.5$\,\lup}{[19.6, 27.4]} & \cv{61.7}{$+20.9$\,\lup}{[19.2, 22.5]} & \cv{55.6}{$+23.3$\,\lup}{[21.4, 25.2]} & \cv{40.1}{$+24.4$\,\lup}{[20.5, 29.0]} & \cv{47.5}{$+29.5$\,\lup}{[26.1, 32.8]} \\
\midrule
\multicolumn{9}{l}{\textit{Transcription systems, used outside their purpose (note output, no added threshold)}} \\
MuScriptor CFG2 & MIDI synth., internal & \cv{77.1}{$-0.1$\,\lne}{[$-2.8$, 2.3]} & \cv{76.2}{$+1.4$\,\lne}{[$-2.3$, 5.2]} & \cv{75.5}{$+1.7$\,\lne}{[$-0.1$, 3.4]} & \cv{\textbf{85.9}}{$-3.3$\,\ldn}{[$-4.7$, $-2.0$]} & \cv{79.1}{$-0.2$\,\lne}{[$-2.8$, 3.2]} & \cv{71.3}{$-6.8$\,\ldn}{[$-9.1$, $-4.4$]} & \cv{77.2}{$-0.2$\,\lne}{[$-2.4$, 2.5]} \\[5pt]
YourMT3+ & incl.\ URMP, MusicNet & \cv{\textbf{77.5$^\dagger$}}{$-0.5$\,\lne}{[$-3.8$, 2.2]} & \cv{\textbf{78.0}}{$-0.3$\,\lne}{[$-8.2$, 8.7]} & \cv{55.5}{$+21.6$}{[17.4, 25.9]} & \cv{72.5}{$+10.1$}{[7.7, 12.5]} & \cv{55.4}{$+23.5$}{[18.2, 28.5]} & \cv{51.3$^\S$}{$+13.2$}{[9.9, 16.9]} & \cv{52.2$^\S$}{$+24.8$}{[19.8, 29.5]} \\
\bottomrule
\end{tabular}}
\caption{Comparison on the semitone metric. Each cell gives the recording-level macro F1 (\%) in normal type, the highest in each column in bold, and below it in small type the paired difference $\Delta$ (points) with its 95\% bootstrap CI. $\Delta$ = JIT-MPE $-$ system, so positive values favour JIT-MPE; JIT-MPE is read at $D=50$\,ms (stale hold), and the systems are scored at full length. \lup\ above, \lne\ not significantly different, \ldn\ below. For the transcription systems only \lne\ and \ldn\ are marked. Column headers give recordings/work clusters. $\ast$: uniform check only; the primary truth of ChoraleBricks is f0 (Table~\ref{tab:f0}). $\ddagger$: partly in-domain for JIT-MPE through MusicNet recordings (note-19) or repertoire (BSED). $\dagger$: system trained on recordings from the same dataset as part of the column. $\S$: system trained on MusicNet, whose repertoire overlaps BSED. Bounds that would round to zero are given to their first nonzero decimal. MAE/Slakh/GS/iK/MDB: the Basic Pitch training data of Table~\ref{tab:f0}.}
\label{tab:semi}
\end{table*}

\subsection{Misses after note onsets}\label{sec:onset}

\begin{figure}[!tb]
  \centering
  \includegraphics[width=\columnwidth]{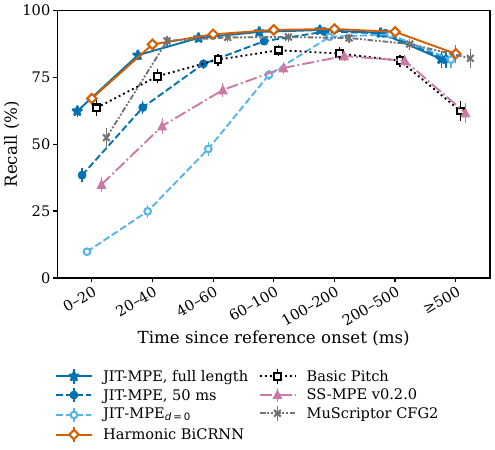}
  \caption{Recall by time since onset on held-out AAM. Each reference frame is assigned to a bin by the time since the onset of its pitch. JIT-MPE is read at $D = 50$\,ms, at full length (160\,ms lookahead) and through its $d=0$ head alone; the other systems are scored at full length. Error bars: 95\% CIs.}
  \label{fig:onset}
\end{figure}

Where does the AAM cost come from? Figure~\ref{fig:onset} splits recall by the time since the onset of each reference pitch. In the first 20\,ms after an onset, JIT-MPE recalls 38.4\% of the reference frames at 50\,ms and 62.3\% with 160\,ms of lookahead; \jitz\ recalls 9.8\% and Harmonic BiCRNN 67.0\%. The difference between 50\,ms and full length shrinks to 19.5, 9.7 and 3.5 points in the next three bins and is below 1 point from 100\,ms on. Of the 13.6k reference frames that JIT-MPE additionally misses at 50\,ms on AAM (net, mean over seeds), 79\% lie within 60\,ms of an onset; on BSED-real and BSED-synthetic the share is 50\% and 57\%. The additional false positives (17.5k on AAM) are not concentrated near onsets, and micro precision falls from 83.4\% to 82.0\%. The difference between full-length JIT-MPE and Harmonic BiCRNN lies elsewhere: 16.9k of the 20.3k frames that JIT-MPE additionally misses there are 500\,ms or more after an onset. In the first bin, the plotted published systems recall 34.9\% (SS-MPE v0.2.0), 52.4\% (MuScriptor CFG2) and 63.6\% (Basic Pitch) at full length, and the unplotted WP-DRCNN and YourMT3+ 70.7\% and 74.2\%.

\begin{table*}[t]
\centering
{\footnotesize\setlength{\tabcolsep}{5pt}\begin{tabular}{lccccc}
\toprule
 & Development & \multicolumn{4}{c}{Held-out} \\
\cmidrule(lr){2-2}\cmidrule(lr){3-6}
Model & All-29 & AAM & ChoraleBricks & BSED-real & BSED-syn. \\
\midrule
JIT-MPE & \textbf{80.83} & \textbf{82.58} & \textbf{77.96} & \textbf{64.52} & \textbf{77.01} \\[6pt]
\quad$-$ later read of the delay heads (= \jitz) & \cv{77.24}{$+3.59$\,\lup}{[2.81, 4.51]} & \cv{80.41}{$+2.17$\,\lup}{[1.96, 2.40]} & \cv{77.95}{$+0.02$\,\lne}{[$-0.09$, 0.12]} & \cv{62.81}{$+1.70$\,\lup}{[1.33, 2.05]} & \cv{74.32}{$+2.69$\,\lup}{[2.32, 2.97]} \\[6pt]
\quad$-$ window and multi-head training (= Harmonic CRNN) & \cv{75.69}{$+1.55$\,\lup}{[1.09, 2.08]} & \cv{80.32}{$+0.09$\,\lne}{[$-0.28$, 0.44]} & \cv{77.66}{$+0.28$\,\lne}{[$-0.24$, 0.80]} & \cv{62.99}{$-0.18$\,\lne}{[$-0.50$, 0.14]} & \cv{73.31}{$+1.01$\,\lup}{[0.56, 1.53]} \\[6pt]
\quad$-$ harmonic front end (= Dilated CRNN) & \cv{75.13}{$+0.55$\,\lup}{[0.18, 0.91]} & \cv{80.34}{$-0.02$\,\lne}{[$-0.46$, 0.44]} & \cv{77.92}{$-0.26$\,\lne}{[$-0.76$, 0.24]} & \cv{60.78}{$+2.20$\,\lup}{[1.86, 2.62]} & \cv{72.17}{$+1.14$\,\lup}{[0.02, 2.11]} \\
\bottomrule
\end{tabular}}
\caption{Ablation of JIT-MPE at $D = 50$\,ms on the stale-hold clock, primary truth; all rows are our causal models. Each row removes one step from the model of the row above, so the single-head models give their $d = 0$ output. Each cell gives the recording-level macro F1 (\%) in normal type, the highest in each column in bold, and below it in small type the paired difference $\Delta$ (points) with its 95\% bootstrap CI. $\Delta$ = row above $-$ this row, the gain of the removed step. \lup\ above, \lne\ not significantly different, \ldn\ below. Bounds that would round to zero are given to their first nonzero decimal.}
\label{tab:ladder}
\end{table*}

\subsection{Comparison with published systems}\label{sec:published}

Tables~\ref{tab:f0} and~\ref{tab:semi} compare JIT-MPE at 50\,ms with the published systems at full length. They place the absolute accuracy. On the primary truth of each set, JIT-MPE is significantly higher than five of the six MPE systems in 38 of their 45 comparisons; it is lower in two, both on BSED-real, and not significantly different in the other five. On the held-out sets alone, it is higher in 15 of 20; of all 38, 18 are against SS-MPE v0.1.0 and Timbre-Trap.

\textbf{Continuous-f0 truth.} On URMP and Bach10 (Table~\ref{tab:f0}), JIT-MPE is above Basic Pitch, both WP models, Timbre-Trap and SS-MPE v0.1.0, and below SS-MPE v0.2.0. On held-out ChoraleBricks, it is not significantly different from Basic Pitch and WP-DRCNN, above WP-PUnet XL, SS-MPE v0.1.0 and Timbre-Trap, and below SS-MPE v0.2.0. SS-MPE v0.2.0 is thus higher on all three sets, including the two outside its training data.

\textbf{Note truth.} On the development sets of Table~\ref{tab:semi}, JIT-MPE is above Basic Pitch, Timbre-Trap and SS-MPE v0.1.0. Against SS-MPE v0.2.0 and both WP models, the difference is not significant on note-19, where all three are in-domain through its URMP or MusicNet recordings. On TRIOS, JIT-MPE is not significantly different from SS-MPE v0.2.0 and above both WP models; on PHENICX, which has only four pieces, it is above all three. On held-out AAM, JIT-MPE is above all six MPE systems, and on BSED-synthetic it is above five and not significantly different from WP-DRCNN. In the uniform semitone check on ChoraleBricks, WP-DRCNN is narrowly higher ($-1.4$ [$-2.9$, $-0.001$]), whereas on the primary f0 truth the difference is not significant.

\textbf{BSED-real.} This set is the main exception. JIT-MPE reaches 64.5\% F1 there, its lowest value on any set, and is below WP-DRCNN ($-7.9$ [$-9.9$, $-6.0$]), WP-PUnet XL ($-2.5$ [$-4.5$, $-0.3$]) and SS-MPE v0.2.0 ($-2.0$ [$-3.6$, $-0.5$]). The deadline does not explain this: at full length JIT-MPE reaches 65.5\% and Harmonic BiCRNN 65.0\%, against 72.4\% for WP-DRCNN. Both WP models are trained on MusicNet, whose repertoire overlaps BSED, but this overlap alone does not explain the gap either: BSED-synthetic renders the same 20 excerpts, and there JIT-MPE is not significantly different from WP-DRCNN ($-1.0$ [$-2.3$, 0.3]) and above WP-PUnet XL ($+3.5$ [1.4, 5.6]).

\textbf{Transcription systems.} Against MuScriptor CFG2, the difference is not significant on the three development sets, ChoraleBricks and BSED-synthetic. MuScriptor is higher on held-out AAM ($-3.3$ [$-4.7$, $-2.0$]) and BSED-real ($-6.8$ [$-9.1$, $-4.4$]), with full-length F1 of 85.9\% and 71.3\% against 82.6\% and 64.5\% for JIT-MPE at 50\,ms. Against YourMT3+, the difference is not significant on note-19 and TRIOS, and its frame-level scores are lower than those of JIT-MPE on PHENICX and on all held-out sets.

\begin{table}[t]
\centering
{\footnotesize\setlength{\tabcolsep}{3.5pt}\begin{tabular}{lrrl}
\toprule
System & Parameters & GMAC/s & Output \\
\midrule
\multicolumn{4}{l}{\textit{Ours}} \\
JIT-MPE & 86k & 0.68 & causal, $D = 50$\,ms \\
Harmonic BiCRNN & 84k & 0.66 & offline \\
\specialrule{.4pt}{\aboverulesep}{1.5pt}\specialrule{.4pt}{0pt}{\belowrulesep}
\multicolumn{4}{l}{\textit{Published MPE systems}} \\
Basic Pitch & 17k & 0.30 & offline \\
SS-MPE v0.1.0 & 545k & 0.65 & offline \\
SS-MPE v0.2.0 & 1.30M & 2.5 & offline \\
Timbre-Trap & 614k & 7.0 & offline \\
WP-DRCNN & 1.60M & 1{,}047$^a$ & offline \\
WP-PUnet XL & 14.6M & 1{,}762$^a$ & offline \\
\midrule
\multicolumn{4}{l}{\textit{Published transcription systems}} \\
YourMT3+ & 47.1M & 49$^b$ & offline \\
MuScriptor CFG2 & 1.37B & 357$^b$ & offline \\
\bottomrule
\end{tabular}}
\caption{Model size and computation. Parameters: trained weights, without the fixed time-frequency front end (spectrogram, CQT, HCQT or sliCQ) and, for our models, without the auxiliary onset layer. GMAC/s: multiply-accumulates per second of audio, as executed by each system's released inference code (ours: our implementation) on the first 20\,s of a held-out ChoraleBricks recording, counted with the PyTorch FLOP counter (Basic Pitch: from its ONNX graph), front end excluded. Output: causal at the 50\,ms deadline, or offline, computed with the whole recording available. $^a$\,The released code recomputes a 75-frame context for every output frame. $^b$\,Autoregressive decoding, so the count grows with the number of notes; with guidance, MuScriptor runs its decoder twice.}
\label{tab:footprint}
\end{table}

\subsection{Ablation}\label{sec:ablation}

\textbf{The delay heads give the largest gain.} Table~\ref{tab:ladder} removes the three steps of Section~\ref{sec:variants} from JIT-MPE one at a time, in the order of their gain on the development data. Reading the delay heads at 50\,ms instead of at once (JIT-MPE $-$ \jitz) adds 3.59 points on the development data and 2.17, 1.70 and 2.69 points on AAM, BSED-real and BSED-synthetic; on ChoraleBricks it is not significant ($+0.02$). This is the step specific to the deadline, and on the development data, AAM and BSED-synthetic it is larger than the other two steps together.

\textbf{Window and multi-head training.} The asymmetric windows together with multi-head training (\jitz\ $-$ Harmonic CRNN) add 1.55 points on the development data; on the held-out sets they are significant only on BSED-synthetic ($+1.01$). The immediate estimate of JIT-MPE is thus on no set significantly lower than the output of the single-head model.

\textbf{Harmonic front end.} The harmonic front end (Harmonic $-$ Dilated CRNN) adds 0.55 points on the development data and is not significant on AAM and ChoraleBricks. It is significant on both BSED sets, and on BSED-real it is the largest of the three steps ($+2.20$, against $+1.70$ for the delay heads).

\subsection{Size and computation time}\label{sec:compute}

Table~\ref{tab:footprint} gives the size and computation of every system. JIT-MPE has fewer parameters than every published system except Basic Pitch (17k): SS-MPE, Timbre-Trap and WP-DRCNN have 6 to 19 times as many, WP-PUnet XL 170 times and the two transcription systems about 550 and 16,000 times. Its 0.68\,GMAC per second of audio is likewise below that of every published system except Basic Pitch (0.30) and SS-MPE v0.1.0 (0.65); JIT-MPE is significantly more accurate than both on every primary-truth set except ChoraleBricks, where it is not significantly different from Basic Pitch (Section~\ref{sec:published}). About a third of the count of JIT-MPE (0.24\,GMAC/s) is multiplication by the zero padding of the harmonic gather in our implementation. Its advantage over the larger systems thus does not come from size.

We timed JIT-MPE on one thread of an Intel Core i9-13900KF CPU, from the update of the audio history through the front end, stem and all five presence heads to the calibrated threshold, over 3,000 hops of synthetic three-tone audio after 100 warm-up hops, excluding the 20-cent refinement used only for continuous-pitch scoring. The median time per 20\,ms hop is 2.42\,ms (2.416--2.434\,ms across seeds), the 95th percentile 2.70--2.80\,ms and the maximum 3.86--4.15\,ms. At $D = 50$\,ms the selected head leaves at least 10\,ms between the arrival of the audio it needs and the deadline (Section~\ref{sec:protocol}), so even the slowest measured hop completes in time.

\section{Discussion}\label{sec:discussion}

\subsection{Why waiting helps after onsets}

When a note starts, the windows of the $d = 0$ head contain at most a few milliseconds of it, and the onset transient is not yet resolved into a pitch. In the first 20\,ms after an onset on AAM, recall rises from 9.8\% for the $d = 0$ head to 38.4\% for the head read at 50\,ms and 62.3\% with 160\,ms of lookahead, close to the 67.0\% of Harmonic BiCRNN (Section~\ref{sec:onset}). On AAM, the frames that the deadline additionally misses are thus mostly those of notes that have just started, whereas the misses that remain against the bidirectional model lie in sustained notes.

This pattern is what the delay heads address. JIT-MPE gives an immediate estimate for every 20\,ms frame and revises that frame as further audio arrives, up to 160\,ms later. Compared with withholding the output until the final estimate, this gives a first answer up to 160\,ms sooner; compared with keeping only the immediate estimate, it also delivers the accuracy that the later updates add, which on AAM lies mostly at note starts.

\subsection{Limitations}

Our results concern frame-level pitch presence. Note-level transcription, singing voice and real recordings of popular music are not evaluated, nor is how a listener perceives a display that rewrites its recent history. We have not trained a model for one fixed 40\,ms lookahead and do not claim that the delay heads are more accurate than such a model. Appendix~\ref{s:caveats} discusses further caveats of our analysis.

\section{Conclusion}\label{sec:conclusion}

JIT-MPE, a causal MPE model of 86k parameters for mixed ensembles, gives an immediate estimate of each frame and re-estimates it as more audio arrives. Scored only on outputs that can be computed from the audio available by the deadline, its update at a 50\,ms deadline, made with 40\,ms of lookahead, is significantly higher than five of six published offline MPE systems in 38 of 45 comparisons and lower only in two, although these systems read the whole recording and all of them except Basic Pitch are larger. The deadline costs little: on the four held-out sets, JIT-MPE at 50\,ms is within $-1.27$ to $+0.60$ points of recording-level macro F1 of its own final output with 160\,ms of lookahead and within $-2.08$ to $+0.79$ points of a bidirectional model of similar size. Waiting improves above all what the immediate estimate misses just after note onsets: on AAM, the recall of reference frames in the first 20\,ms after an onset rises from 9.8\% to 38.4\% at 50\,ms and 62.3\% at 160\,ms, and the update at 50\,ms adds 1.7--3.6 points over the immediate estimate on the development data and on three of the four held-out sets; on ChoraleBricks, waiting does not help. Larger models, note-level output and recordings closer to live performance are the natural next steps.

\section*{Code Availability}

Training and evaluation code and the trained weights of our models (JIT-MPE and the six comparison models of Section~\ref{sec:variants} and Appendix~\ref{s:model}) are available under the MIT licence at \url{https://github.com/Frieve-A/jit-mpe}; its README maps the model names of this paper to the identifiers used in the code. The NoteSpectrogram analyzer of the EffeTune DSP Library is derived from JIT-MPE but trained on different data, so the results in this paper do not describe it.

\section*{AI Usage Statement}

Language models assisted with code, experiment orchestration and drafting of the text. All reported numbers were produced by the pipeline described in this paper and checked by the author.

\bibliographystyle{IEEEtran}
\bibliography{refs}

\appendix

\section{Input and Model}\label{s:model}

\textbf{Input.} All audio is processed at 44.1\,kHz with a hop of 882 samples (20\,ms). Each frame uses two windows of 4096/48 and 16384/48\,ms (85.33 and 341.33\,ms) that end at the frame end. A window of duration $T$ spans $\mathrm{round}(T f_s)$ samples, 3763 and 15053 at 44.1\,kHz, and the FFT length is the smallest power of two not below that length, 4096 and 16384. Each window $w$ is divided by its sum, so the magnitude $a = |\mathrm{rfft}(w \cdot x)|$ is normalized by the window sum. The magnitudes are projected onto 300 log-frequency bins with centres $m_k = 20\tfrac{2}{3} + k/3$, $k = 0, \ldots, 299$, in MIDI units. Each bin takes a sum-normalized triangular weighting of FFT bins; where the FFT resolution is too coarse for this, it interpolates between the two nearest FFT bins. The feature is $\log(1 + 10^4 a)$, standardized per window and bin with the mean and standard deviation of training frames (standard deviation floored at 0.01).

\textbf{Windows.} JIT-MPE uses asymmetric windows: a Hann rise over all but the last $F$ samples, followed by a Hann fall over $F$ samples, with $F = \mathrm{round}(t_F f_s)$, $t_F = 10$\,ms for the short window and 20\,ms for the long one. All other models use periodic Hann windows of the same lengths.

\textbf{CRNN front end (Plain and Dilated CRNN).} Three $3 \times 7$ convolutions (time $\times$ frequency) with 32 channels and ReLU, causal in time, with frequency dilations 1, 1, 1 (Plain) or 1, 6, 36 (Dilated), are followed by a projection over 25 bins with stride 3 to 47 features per note. The frequency reach is $3(1+1+1) + 12 = 21$ bins (Plain) or $3(1+6+36) + 12 = 141$ bins (Dilated). A 47-unit GRU shared across notes follows.

\textbf{Narrow front end (Plain CRNN and Plain BiCRNN).} Plain CRNN has 95,767 parameters and Plain BiCRNN, its bidirectional counterpart with 26 units per direction, 93,936. Their F1 on All-29 is 62.81\% and 67.87\%, against 75.93\% and 79.84\% for Dilated CRNN and Dilated BiCRNN; Table~\ref{tab:abs} lists all sets.

\textbf{Harmonic stem (Harmonic CRNN and JIT-MPE).} The stem has $C = 24$ channels and eight layers with $3 \times 7$ kernels, causal in time, so that it sees the last 17 frames. Layer 0 is a convolution from the two window channels to $C$ channels followed by a ReLU. Layers 1 to 7 are depthwise-separable convolutions with frequency dilations 1, 2, 4, 8, 16, 32 and 64, each applied as a residual $h \leftarrow h + \mathrm{ReLU}(y)$. The frequency reach is $3 \times (1 + 1 + 2 + \cdots + 64) = 384$ bins.

\textbf{Harmonic gather.} Note $p = 0, \ldots, 87$ (MIDI $21 + p$) reads the stem features at the bins
\[
k_{p,h} = 3p + 1 + \lfloor 36 \log_2 h + 0.5 \rfloor
\]
for the 18 harmonic ratios $h \in \{0.5, 1, 1.5, 2, 3, 4, \ldots, 16\}$. With the stem features $x$ and weights $W_h$ shared across notes,
\[
v_p = \mathrm{ReLU}\Big(\sum_h W_h\, x[k_{p,h}] + b + e_p\Big),
\]
which gives 64 features per note; $e_p$ is a learned per-note term initialized to zero. Taps outside the 300 bins read zero.

\textbf{Interval aggregation.} With the semitone offsets $O = \{0, \pm 1, \pm 2, \pm 5, \pm 7, \pm 12, \pm 19, \pm 24\}$,
\begin{align*}
u_p &= \mathrm{ReLU}\Big(\sum_{o \in O} U_o v_{p+o} + b_U\Big),\\
v_p &\leftarrow \mathrm{LayerNorm}(v_p + B u_p + b_B),
\end{align*}
where $U_o$ maps 64 to 16 features, $B$ maps 16 to 64 features and only the weight of $B$ is initialized to zero. Offsets $p + o$ outside the 88 notes read zero.

\textbf{Recurrence and heads.} In Harmonic CRNN and JIT-MPE, a GRU with 64 units runs over time for each note, with weights shared across notes. The presence head of delay $d$ of JIT-MPE reads the GRU state of row $t + d$ through a linear map from 64 units to the five delays, shared across notes, plus a per-pitch, per-delay bias initialized to $-3$. The other causal models have $d = 0$ only. All models, including the BiCRNNs, also have onset heads for delays 1, 2 and 4 with per-pitch weights and a bias initialized to $-4$; they serve only the auxiliary loss of stage 2.

\textbf{Bidirectional counterparts.} The BiCRNNs use the same frequency layout with time-centred convolutions and a bidirectional GRU (26 units per direction for Plain and Dilated BiCRNN, 37 for Harmonic BiCRNN). They are trained on the same crops and run on the whole recording.

\textbf{Times and losses.} Row $t$ of a crop has an input window ending at the frame end $e_t$; label row $t$ refers to the label time $e_t + 10$\,ms; and a head of delay $d$ read at row $t$ is compared with label row $t - d$. Labels take the values 0, 1 and 2 (uncertain). The presence loss $L_v$ is the binary cross-entropy of each presence head, with uncertain labels and the first 16 rows of each crop (burn-in) masked, averaged over heads. In stage 2, the onset target of a pitch at label time $s$ is
\[
y_o(s) = \max_c \max\big(0,\, 1 - |s - c| / (2 \cdot 20\,\mathrm{ms})\big)
\]
over the onsets $c$ of that pitch, with zero weight where the presence label is uncertain. The onset loss $L_o$ is a masked binary cross-entropy with positive weight 10, averaged over the three onset heads, and the total loss is $L_v + 0.18\,L_o$. Training uses AdamW with weight decay $10^{-4}$, a one-cycle schedule with 5\% warm-up, batches of 16 crops of 192 frames (3.84\,s) and gradient-norm clipping at 1. The peak learning rate is 0.002 in stage 1 and 0.0005 in stage 2.

\section{Training Data and Labels}\label{s:data}

\begin{table}[t]
\centering
\small
\setlength{\tabcolsep}{3pt}
\begin{tabular}{lrl}
\toprule
Dataset & Clips & Label source \\
\midrule
GuitarSet & 138 & note annotations \\
Slakh2100-redux & 137 & MIDI notes with sustain pedal \\
MusicNet & 67 & aligned nominal note pitches \\
Winterreise & 10 & aligned nominal note pitches \\
Cantoria & 64 & union of per-voice pYIN tracks \\
Dagstuhl ChoirSet & 9 & aligned score notes \\
vocadito & 29 & human f0 annotations \\
Groove & 313 & none (drums only) \\
\midrule
Total & 767 & \\
\bottomrule
\end{tabular}
\caption{Training clips by dataset. About one song in eight of each dataset is held out from training; these clips, except those of Groove, form the calibration set (Appendix~\ref{s:calib}). Cantoria, Dagstuhl ChoirSet and vocadito form the vocal pool.}
\label{tab:clips}
\end{table}

\textbf{Label rules.} Labels are 88-key semitone presence at the label times. For note labels, a pitch is active at time $t$ if $\mathrm{on} \le t < \mathrm{off}$, and notes of the same pitch are combined by OR. GuitarSet takes $\lfloor m + 0.5 \rfloor$ of the annotated MIDI pitch $m$; MusicNet and Winterreise use aligned nominal pitches and Dagstuhl ChoirSet aligned score notes. Slakh uses the MIDI notes extended by the sustain pedal, with a fixed semitone correction per rendering patch. For the f0 labels of Cantoria and vocadito, the voiced f0 at the nearest time is rounded to the nearest semitone, and Cantoria combines its per-voice tracks by OR. Labels are marked uncertain where they cannot be trusted: frames where a pitch outside MIDI 21--108 sounds, pitch bends and digital silence in Slakh, frames next to note boundaries in Dagstuhl ChoirSet and next to voicing boundaries in the f0 labels.

\textbf{Sampling.} Each step draws 16 random 192-frame crops within the labelled range of the training clips. A two-level sampler first picks the instrumental or the vocal pool in proportion to its labelled frames and then a clip in proportion to its labelled frames, with Groove down-weighted to 0.3.

\textbf{Stage-2 augmentation.} Drum excerpts from Groove and from Slakh drum stems are mixed into crops at random levels, and some Groove crops are replaced by Slakh drum excerpts. Vocal excerpts from the vocal pool are mixed in with their labels. Crops without low notes (none in MIDI 28--47) receive a line of synthetic tones in MIDI 21--60, mostly in the lowest octave, with sine, harmonic or resonator timbres and sustained, slow-attack, short or decaying envelopes, and the tones are added to the labels. Finally, a random compressor with soft clipping and a random level change are applied. Onsets for the auxiliary loss come from the note starts of the note-labelled datasets, from voicing starts in the vocal pool and from the start times of the synthetic tones.

\section{Calibration and Head Selection}\label{s:calib}

\textbf{Calibration set.} The held-out calibration clips of the seven training datasets other than Groove, each used clean and with a fixed mixing that adds a looped drum clip from another song, the compressor of stage 2 and a level change, at equal weight.

\textbf{Bias.} For each run and head, a global bias and three register biases are chosen from $-1.5$ to $1.5$ in steps of 0.05. At inference, pitch $p$ of output row $t$ from head $d$ is active if $z_{t+d,p,d} + b_{n} \ge 0$, where $z$ is the presence logit and $n$ the MIDI note of $p$. The bias $b_n$ is the register bias for $21 \le n \le 47$, $48 \le n \le 71$ and $72 \le n \le 91$, which replaces the global bias there, and the global bias for $92 \le n \le 108$. The global bias is fitted on all 88 notes and each register bias on its own register. The objective pools true positives, false positives and false negatives within each dataset, takes the F1 of each dataset, averages it with equal weight over the seven datasets and then averages the clean and mixed versions; ties go to the smaller bias.

\textbf{Selection set.} One head per deadline is chosen on the stale-hold clock on 187 clips that exclude All-29 and the held-out sets, all disjoint by work from our training clips: GuitarSet (10), MusicNet (91, from its official training split; the MusicNet recordings of All-29 come from its test split), Slakh2100-redux (80, from its official validation split) and Winterreise (6). The criterion is the micro F1 pooled within each dataset, averaged with equal weight over the four datasets. The candidates for deadline $D$ are the heads with $20d \le D$\,ms, and ties go to the smaller $d$. One full-length head is chosen by the same criterion from all heads.

\section{Evaluation Data and Scoring}\label{s:eval}

\begin{table*}[t]
\centering
\small
\begin{tabular}{>{\raggedright}p{2.2cm}>{\raggedright}p{4.5cm}>{\raggedright}p{3.3cm}>{\raggedright\arraybackslash}p{4.1cm}}
\toprule
System & Source & Checkpoint & Training and selection \\
\midrule
SS-MPE v0.1.0 & \path{github.com/cwitkowitz/ss-mpe}, release v0.1.0 & SS-MPE experiment & self-supervised on NSynth, chosen by NSynth validation loss \\[3pt]
SS-MPE v0.2.0 & \path{github.com/cwitkowitz/ss-mpe}, release v0.2.0 & URMP supervised & supervised on URMP, chosen on the URMP F0 recordings \\[3pt]
Basic Pitch & \path{github.com/spotify/basic-pitch} & ICASSP 2022 model & MAESTRO, Slakh, GuitarSet, iKala, MedleyDB \\[3pt]
Timbre-Trap & authors' release & original model & URMP, selected on a nine-recording URMP subset \\[3pt]
WP-DRCNN & \path{github.com/christofw/}\newline\path{multipitch_architectures} & DRCNN size L, MuN-10full & MusicNet \\[3pt]
WP-PUnet XL & \path{github.com/christofw/}\newline\path{multipitch_architectures} & PUnet XL, MuN-10full & MusicNet \\[3pt]
YourMT3+ & authors' release & YPTF.MoE+Multi (noPS) & includes URMP and MusicNet \\[3pt]
MuScriptor & authors' release & muscriptor-large, guidance strength 2 (CFG2) & audio synthesized from public and commercial MIDI, internal real recordings \\
\bottomrule
\end{tabular}
\caption{Published systems. All are offline and are scored on their full-length outputs. Their size and computation are given in Table~\ref{tab:footprint}.}
\label{t:ext}
\end{table*}

\begin{table*}[t]
\centering
{\footnotesize\setlength{\tabcolsep}{4pt}\begin{tabular}{lccccccccc}
\toprule
 & \multicolumn{5}{c}{Development} & \multicolumn{4}{c}{Held-out} \\
\cmidrule(lr){2-6}\cmidrule(lr){7-10}
System & URMP f0 & Bach10 f0 & note-19 & TRIOS & PHENICX & AAM & ChoraleBricks f0 & BSED-real & BSED-syn. \\
\midrule
\multicolumn{10}{l}{\textit{Our models (causal models at $D = 50$\,ms unless noted; bidirectional models at full length)}} \\
JIT-MPE, 50\,ms & 84.7 & 91.1 & 77.0 & 77.6 & 77.2 & 82.6 & 78.0 & 64.5 & 77.0 \\
JIT-MPE, full length & 83.4 & 90.0 & 77.0 & 77.3 & 77.2 & 83.9 & 77.4 & 65.5 & 78.3 \\
\jitz & 82.3 & 88.3 & 73.0 & 75.6 & 75.7 & 80.4 & 77.9 & 62.8 & 74.3 \\
Harmonic CRNN & 81.2 & 86.9 & 71.5 & 75.1 & 75.4 & 80.3 & 77.7 & 63.0 & 73.3 \\
Dilated CRNN & 80.8 & 86.8 & 70.7 & 72.7 & 74.8 & 80.3 & 77.9 & 60.8 & 72.2 \\
Plain CRNN & 65.5 & 74.6 & 57.0 & 62.3 & 61.4 & 71.0 & 69.0 & 46.8 & 54.4 \\
Harmonic BiCRNN & 83.6 & 90.0 & 77.1 & 78.3 & 78.5 & 84.7 & 77.2 & 65.0 & 77.5 \\
Dilated BiCRNN & 83.0 & 89.5 & 76.4 & 77.1 & 78.8 & 83.7 & 78.5 & 64.2 & 75.7 \\
Plain BiCRNN & 69.4 & 79.2 & 63.3 & 64.5 & 65.2 & 76.0 & 67.8 & 48.3 & 60.1 \\
\midrule
\multicolumn{10}{l}{\textit{Published systems (full length)}} \\
SS-MPE v0.1.0 & 60.8 & 72.3 & 54.8 & 58.0 & 54.0 & 60.9 & 69.8 & 45.5 & 51.9 \\
SS-MPE v0.2.0 & 88.4 & 92.9 & 75.0 & 72.7 & 74.3 & 75.6 & 80.9 & 66.6 & 70.6 \\
Basic Pitch & 80.9 & 88.3 & 72.6 & 69.4 & 74.2 & 74.1 & 78.0 & 57.9 & 65.9 \\
WP-DRCNN & 76.6 & 82.3 & 76.4 & 71.6 & 73.1 & 76.1 & 78.2 & 72.4 & 78.0 \\
WP-PUnet XL & 74.7 & 82.2 & 74.8 & 66.7 & 65.7 & 72.4 & 73.4 & 67.0 & 73.6 \\
Timbre-Trap & 73.3 & 84.7 & 62.1 & 58.1 & 53.7 & 61.7 & 54.8 & 40.1 & 47.5 \\
MuScriptor CFG2 & 80.7 & 79.0 & 77.1 & 76.2 & 75.5 & 85.9 & 76.2 & 71.3 & 77.2 \\
YourMT3+ & 81.7 & 48.1 & 77.5 & 78.0 & 55.5 & 72.5 & 54.1 & 51.3 & 52.2 \\
\bottomrule
\end{tabular}}
\caption{Absolute recording-level macro F1 (\%) on the primary truth of each set: continuous-f0 F1 for URMP, Bach10-mf0-synth and ChoraleBricks, and semitone F1 otherwise. Our models: mean of three seeds; outputs at 50\,ms on the stale-hold clock, full-length outputs on the nearest output frame (Section~\ref{sec:metrics}).}
\label{tab:abs}
\end{table*}

\textbf{AAM.} From songs 1 to 2000 of the 3,000 in the collection, reserved before development, we take the first 100 songs of a fixed pseudo-random order defined in the released code; no content-based selection is made. Each excerpt covers $[30, 60)$\,s, or the last 30\,s for songs shorter than 60\,s, and is down-mixed to mono by channel mean. The truth is the nominal pitch of the per-instrument MIDI notes, active for $\mathrm{on} \le t < \mathrm{off}$; drum notes are excluded, and pitches outside MIDI 21--108 are marked uncertain.

\textbf{ChoraleBricks (v1.1.0).} Each mixture is the equal-weight average of all distributed normalized stems of the song, aligned at a common origin and cut to the length of the shortest stem. The continuous truth uses the recommended filled f0 track of each stem, resampled to the 10\,ms query grid by nearest neighbour and kept only where voiced; values from different stems within the same MIDI bin are merged by geometric mean. The semitone truth uses the nominal notes, and the offset $c$ uses all voiced f0 values.

\textbf{BSED (v1.0.0).} We use the 44.1\,kHz audio and the note-level alignment MIDI. Note start and end times are rounded up to the next 10\,ms query time, and a note is active for $\mathrm{on} \le t < \mathrm{off}$. Notes of the same pitch are combined by OR, drums are excluded, and pitch bends and pedal are not applied. The mixture is the channel mean, and the truth is the nominal note pitch with $c = 0$.

\textbf{Published systems.} Table~\ref{t:ext} lists the checkpoints. All thresholds are author defaults: SS-MPE and Timbre-Trap use spectral peaks of at least 0.5 up to 5\,kHz, both WP models 0.4, and Basic Pitch 0.3 on its frame output. The transcription systems use their native note output.

\textbf{Absolute scores.} Table~\ref{tab:abs} lists the absolute recording-level macro F1 of all systems on the primary truth of every set.

\textbf{Scoring.} Every system receives 44.1\,kHz mono audio followed by 1\,s of silence. The URMP recordings (in URMP F0s and note-19) are resampled from the 16\,kHz release prepared for YourMT3~\cite{chang2024yourmt3} rather than the 48\,kHz original, so for every system they contain no content above 8\,kHz. All systems are queried at the times of Section~\ref{sec:metrics}.
\begin{enumerate}
\item \emph{Pitch of our models.} Each active note $n$ of an output row is emitted at $440 \cdot 2^{(n + f/5 - 69)/12}$\,Hz with a fine offset $f \in \{-2, \ldots, 2\}$ in 20-cent steps. For $f$, each harmonic $h = 1, \ldots, 4$ of the note below $0.45 f_s$ is searched in the long-window magnitude spectrum of the target row within $\pm 60$\,cents; the largest bin is used if it is a local maximum, refined by quadratic interpolation of the log power. The magnitude-weighted mean deviation $\delta$ in cents gives $f = \min(\max(\lfloor(\delta + 50)/20\rfloor, 0), 4) - 2$. A row whose short-window RMS is below $-96$\,dBFS emits nothing. For $d > 0$, the fine offsets and this silence floor are taken from the target row.
\item \emph{Rows.} Full-length outputs take the output row nearest to each query time (for our models, by target time; an exact tie goes to the earlier row) and are empty outside their range. A streamed output of our models for query time $q$ and deadline $D$ reads row $r$, the latest row whose target time $(r + 1) \cdot 20$\,ms is at most $q$, through the head selected for $D$.
\item \emph{Uncertain cells.} Uncertain truth cells are ignored for both truth and estimates.
\item \emph{Matching.} Continuous-f0 F1 matches estimates and references one-to-one within 50 cents; estimates of the published systems outside MIDI 21--108 count as false positives. The semitone F1 applies the tuning offset $c$ and rounding of Section~\ref{sec:metrics} to every system, restricts all pitches to MIDI 21--108 and merges hits on the same pitch. In the All-29 aggregate, the MusicNet notes are matched like continuous f0, within 50 cents of their nominal pitch; this agrees with their semitone F1 to within 0.01 points except for the continuous outputs of SS-MPE v0.1.0 and Timbre-Trap, which score up to 1 point lower.
\end{enumerate}

\textbf{Bootstrap.} 10,000 draws resample work clusters within each dataset: pieces for the development sets, songs for AAM and ChoraleBricks and symphonies for BSED. Within each draw the three seeds are averaged, and all systems and paired differences use the same resampled recordings; the CIs therefore do not include the variation between training runs.

\section{Scope and Caveats}\label{s:caveats}

\textbf{Comparison models.} The comparison with Harmonic BiCRNN holds for models of fewer than 100k parameters trained on our data; a larger bidirectional model may use future context better and widen the gap.

\textbf{Few clusters.} ChoraleBricks has 10 songs, BSED nine symphonies, TRIOS five and PHENICX four pieces, so their CIs rest on few clusters; 10 of the 38 higher comparisons with published systems come from TRIOS and PHENICX.

\textbf{ChoraleBricks.} On this set the delay heads do not help, and the F1 of JIT-MPE at 50\,ms is above that at full length ($+0.60$), also against the nominal notes; we have not identified the cause. The upper ends of both ranges ($+0.60$ and $+0.79$) come from this set and are comparable to the seed standard deviation there (0.52 points for Harmonic BiCRNN).

\textbf{Output clock.} On URMP and Bach10, whose query grid is offset from our rows, the nearest-row read used for full-length outputs often takes the next row (Section~\ref{sec:metrics}). This lowers the full-length scores of our models there: on the development aggregate, the 160\,ms output scores 80.55\% on the nearest row and 81.03\% on the stale-hold clock, 0.20 points above the 50\,ms read (80.83\%). It also changes the ablation step of the delay heads on the development data, which is $+2.61$ instead of $+3.59$ when each query reads the nearest row, because the $d = 0$ output of a row after $q$ is not yet ready. On the held-out grids the two reads coincide.

\textbf{Onset analysis.} The onset times of AAM are those of the MIDI notes from which it is rendered, so the first bin of Figure~\ref{fig:onset} also includes the attack of each note, during which its pitch may not yet be established. The figure shows the recall of reference frames; it does not measure how late a note is first detected.

\end{document}